\documentclass{llncs}
\usepackage{graphicx, amssymb, epsfig, amsmath}
\usepackage{float, color}

\begin{document}
\mainmatter

\newcommand{\dima}[1]{\textcolor{green}{\bf #1}}
\newcommand{\sean}[1]{\textcolor{blue}{\bf #1}}
\newcommand{\ToDo}[1]{\textcolor{red}{\bf #1}}

\title{Weighted hierarchical alignment of \\ directed acyclic graphs}
\titlerunning{DAG alignment}  
\author{Sean M. Falconer\inst{1} \and Dmitri Maslov\inst{2}}
\authorrunning{Sean M. Falconer et al.}   
\tocauthor{Sean M. Falconer (University of Victoria), Dmitri Maslov
(University of Waterloo)}
\institute{University of Victoria, Victoria, BC, V8W 3P6, Canada,\\
\email{seanf@uvic.ca}
\and
University of Waterloo, Waterloo, ON, N2L 3G1, Canada}

\def\mw#1{\mbox{\em #1}\,}
\def\thru{ \!\! \rightarrow\ \!\!\!\!}
\def\nw#1{\mbox{#1}}

\maketitle

\begin{abstract}
    In some applications of matching, the structural or hierarchical properties of the two graphs being aligned must be maintained.  The hierarchical 
properties are induced by the direction of the edges in the
two directed graphs.  These structural relationships defined by the hierarchy in the graphs act as a constraint on the alignment.  In this paper, we 
formalize the above problem as the weighted alignment between two directed acyclic graphs.  We prove that this problem is NP--complete, show several 
upper bounds for approximating the solution, and finally introduce polynomial time algorithms for sub--classes of directed acyclic graphs.
\end{abstract}

\section{The problem}
\begin{figure}
        \centerline{\includegraphics[scale=0.27]{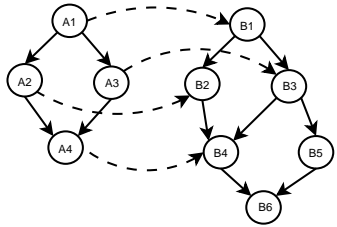}}
        \scriptsize{\caption{Example of a valid matching between two graphs.  The dashed lines represent valid assignments for the vertices $A1, A2, 
A3,$ and $A4$.}}
        \label{fig:valid_matching}
    \end{figure}

Matching or alignment problems are an important set of theoretical problems that appear in many different applications 
\cite{buss95bipartite,caseau97solving,galil86weightedmatching}.  Depending on the structure of the problem, polynomial time algorithms may or may not 
exist.  In this paper, we propose a new type matching problem called the {\it weighted hierarchical DAG (directed acyclic graph) alignment problem}.  
In this problem, we 
have two directed acyclic graphs and a set of possible matchings between vertices in both graphs.  We wish to find the maximum weighted matching 
between the vertices where the directed edges in both graphs act as {\it hierarchical constraints} on possible solutions to the matching.  For 
example, if a vertex $v_1$ has a directed edge to a vertex $v_2$, then any matched vertex to $v_2$ cannot be an ancestor of $v_1$'s matched vertex 
(see Figures \ref{fig:valid_matching} and 2).

We became interested in this problem through our interest in ontology alignment. An ontology is a conceptualization of a domain 
\cite{gruber93transapproach}.  This conceptualization consists of a set of terms with certain semantics and relationships \cite{russel95modern}.  
Generally, the terms are related by $\verb"is_a"$ relationships.  The relationships (edges) and terms (vertices) can be represented as a DAG.  With 
ontology alignment, one wants to align terms from two different ontologies in order to merge, compare, or map the ontologies.  Since the edges of the 
DAG represent an $\verb"is_a"$ relationship, then if we apply the strictist sense of this relationship, it constrains the number of valid matchings, 
because we do not wish to violate this relationship in the corresponding matching.

This type of hierarchical or structural constraint is important in other applications as well.  The domains of SVG ({\it Scalable Vector Graphics}) 
version comparison, source code comparison/merging, UML difference calculation, and file/folder merging, are all instances of hierarchical based 
matching.  For example, an SVG document is rich with structure.  The document defines graphical objects, and how they relate, a form of the 
$\verb"is_a"$ relationship exists through the document graphic layers.  In object--oriented programming, $\verb"is_a"$ relationships exist through the 
definitions of inheritance, and other relationships exist
via class membership.  Similarly, UML diagrams
have structural relationships, and different versions of diagrams sometimes need to be merged or have their differences calculated
for visual comparison \cite{niere04uml}.
Finally, in a file system, the folders represent an embedded hierarchy.

    \begin{figure}
        \centerline{\includegraphics[scale=0.27]{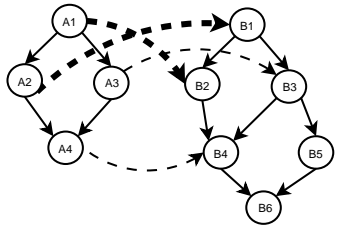}}
        \scriptsize{\caption{Example of an invalid matching between two graphs.  The dashed lines represent the assignments for the vertices $A1, A2, 
A3,$ and $A4$.  The two bold dashed lines represent an assignment violation because $A1$ maps to a descendant of $A2$'s mapped vertex $B1$.}}
        \label{fig:invalid_matching}
    \end{figure}

    \subsection{Related work}
General graph matching is a well studied problem.  Most graph matching problems can be divided into two categories, graph isomorphisms and weighted 
graph matching.  In graph isomorphism, the goal is to find a matching function $f$ for two graphs $G_1 = (V_1, E_1)$ and $G_2 = (V_2, E_2)$.  General 
graph isomorphism is still open, that is, it is not known whether the problem is NP--hard or can be solved in polynomial time 
\cite{garey79npcomplete}.  Sub--graph isomorphism is known to be NP--complete \cite{gold96graduated}.  With weighted graph matching, we are given a 
graph $G = (V, E)$, where the edges have associated weights and we wish to find a subset $M$ of $E$, such that no two edges in $M$ share a common end 
vertex and such that the sum of edge weights in $M$ is maximum.  For some classes of graphs, polynomial time algorithms are known, while some others 
are known to be NP--complete.

Both of these problems have many practical applications, in particular, graph isomorphism has received a lot of attention in the area of computer 
vision.  Images or objects can be represented as a graph.  A weighted graph can be used to formulate a structural description of an object 
\cite{shapiro81inexactmatching}.  There have been two main approaches to solving graph isomorphism: state--space construction with searching and 
nonlinear optimization.  The first method consists of building the state--space, which can then be searched.  This method has an exponential running 
time in the worst case scenario, but by employing heuristics, the search can be reduced to a low--order polynomial for many types of graphs 
\cite{eshera84imageanalysis,tsai83patternrecognition}.  With the second approach (nonlinear optimization), the most successful approaches have been 
relaxation labeling \cite{hummel83relaxation}, neural networks \cite{kuner88neural}, linear programming \cite{almohamad93linear}, eigendecomposition 
\cite{umeyama88eigen}, genetic algorithms \cite{krcmar94genetic}, and Lagrangian relaxation \cite{rangarajan94lagrangian}.

Another type of graph problem related to ours is graph alignment through minimizing the edit distance \cite{zhang95editing,cs.DS/0604037}.
In this problem, the graphs are transformed via editing (deletion, insertion, relabelling) to achieve alignment. Our work is different is several 
ways.  First, we do not allow any of the graph to be edited as is typically done in the edit distance problem. Second, in the work discussed in 
\cite{zhang95editing}, the authors consider only undirected graphs as opposed to DAGs.  Finally, the authors of \cite{cs.DS/0604037} deal with 
unweighted alignment of trees as opposed to weighted alignment of DAGs.

As mentioned, we became interested in DAG alignment problem due to our interests in ontology alignment.  Ontology alignment has recently received a 
lot of attention.  An alignment between two ontologies can be formalized in terms of weighted graph matching, with certain constraints on the solution 
to any valid matching.  Originally, alignments were performed by hand, and later, several researchers introduced semi--automatic alignment strategies, 
which make suggestions to the user about which terms to align \cite{noy02prompt,noy99smart}.  Since then, fully automatic alignment strategies have 
been explored.  In \cite{jrme04state}, over twenty different tools/algorithms are discussed.  Many of these approaches use heuristics to determine 
term similarities, by first comparing syntactic, semantic, and structural similarities, and then compute matches greedily or via some other local 
optimization technique.

In \cite{feng05conceptual}, graph matching is applied to conceptual system matching for translation.  The work is very similar to ontology alignment, 
however, the authors formalize their problem in terms of any conceptual system rather than restricting the work specifically to an ontological 
formalization of a domain.  They formalize conceptual systems as graphs, and introduce algorithms for matching both unweighted and weighted versions 
of these graphs.

    \subsection{Organization of the paper}
    The remainder of the paper is organized as follows.  The next section introduces notations and definitions
that will be used throughout the paper.  The definitions include the formal description of the problem.  Following this, we
show that the decision version of the problem is NP--complete via a reduction from 3SAT.  Next, we prove two theorems, which yield upper bounds on 
approximating the DAG alignment problem.  After this, we introduce a polynomial time algorithm for trees and discuss its possible modifications.  
Finally, we present some concluding remarks, a short discussion of open problems, and directions for future research.

    \section{Notations and definitions}

    \subsection{Notations}
Before formally defining the DAG alignment problem we must first introduce some definitions.  A DAG is a directed graph, $G = (V, E)$ that contains no 
oriented cycles, where $V$ is a set of vertices and $E$ is a set of edges.  Let $anc(v)$ denote the set of ancestors for any $v \in V$, where an 
ancestor of $v$ is any $a \in V$ such that there exists a directed path from $a$ to $v$.  Let $desc(v)$ denote the set of descendants for any $v \in 
V$, where a descendant of $v$ is any $d \in V$ such that there exists a directed path from $v$ to $d$.  Finally, let $child(v)$ denote the set of 
direct children for any $v \in V$, where a direct child is any $d \in V$ such that there exists a directed edge from $v$ to $d$.

    \subsection{Description of problem}
In this section we formalize the problem of DAG alignment with hierarchy constraints.  Without the hierarchy constraint, the problem reduces to 
weighted bipartite matching, since the edges that represent vertex relationships would be ignored.  As was mentioned, in many practical applications 
these structural relationships cannot be ignored.  Due to these relationships, many solutions that would be valid in weighted bipartite matching are 
invalid.  In fact, we can think of any edge $e$ as having a set of conflicting edges, where a conflict is any edge that would violate a matching 
solution that contained $e$.  We formalize this in the following definition.

    \begin{definition}
An {\bf edge conflict} for edge $e = (a, b, w_e)$, $w_e \in [0,1]$,
        is any edge $d = (f, g, w_d)$, $w_d \in [0,1]$, and $d \neq e$, where one of
        the following conditions applies:
        \begin{enumerate}
        \item $a \in anc(f)$ and $b \not\in anc(g)$.
        \item $a \in desc(f)$ and $b \not\in desc(g)$.
        \item $a = f$.
        \item $b = g$.
        \end{enumerate}
    \end{definition}

The set $conf(e)$ denotes the set of edges that have edge conflicts with edge $e$.
We can now give the formal definition of the DAG alignment problem.

    \begin{definition}
        Given two DAGs, $G_1 = (V_1, E_1)$
        and $G_2 = (V_2, E_2)$, and a set of edges $\beta =
        \{(v_i, v_j, w_t)\}$ for all $v_i \in V_1$, all $v_j
        \in V_2$ and $w_t \in [0, 1]$, the {\bf
        DAG alignment problem} is to find the maximum weight matching, $M \subseteq \beta$,
        such that each vertex in $M$ appears only once and for any
        edge $e \in M$, $conf(e) \cap M = \emptyset$.  We refer
        to this constraint on the matching as the {\it hierarchical
        constraint} for the remainder of this paper.
    \end{definition}

Our definition of the DAG alignment problem uses a complete bipartite
graph of all possible matchings with the set of edges $\beta = \{(v_i, v_j, w_t)\}$
defined for all $v_i \in V_1$ and all $v_j \in V_2$. This may appear to narrow
the set of problems we are trying to solve, however, it does not.
This is because a solution to the problem with an incomplete (some
matchings may be inherently prohibitive) matching graph can be reduced to the
problem with complete bipartite graph through the following
consideration. Take a DAG alignment problem in which not every
node of $G_1$ can potentially be mapped to any node of $G_2$.
Allow all the remaining matchings, but assign zero weights to
them. Solve the DAG alignment problem with the complete set of
possible matchings. Delete all zero weight matchings from the
solution. The result is a solution for the DAG alignment problem with
incomplete set of possible matchings.

\section{Intractability}

The DAG alignment problem defined in the previous section is NP--complete.  Before showing the proof of this, we begin by first defining the decision 
version of the problem.

    \begin{definition}
        We are given two DAGs, $G_1 = (V_1, E_1)$
        and $G_2 = (V_2, E_2)$, and a set of edges $\beta =
        \{(v_i, v_j, w_t)\}$ for all $v_i \in V_1$, $v_j
        \in V_2$ and $w_t \in [0, 1]$.  Let
        $w(A)$, where $A \subseteq \beta$, be the sum of all weights
        $w_t$ defined over all triples $(v_i, v_j, w_t) \in A$.
        Is there a matching $M \subseteq \beta$ with
        weight $w(M) \geq X$ and $|M| \leq Y$ such that each vertex
        in $M$ appears only once and for any
        edge $e \in M$, $conf(e) \cap M = \emptyset$?
    \end{definition}

    \begin{theorem}
    DAG alignment, as introduced in Definition 3, is NP--complete.
    \end{theorem}

    \begin{proof}
    It is easy to see that the decision version of DAG
    alignment is in NP, so this will be omitted.

    We show a reduction of 3SAT to the decision version of the DAG alignment problem.  In
    3SAT we have a finite set of variables, $X = \{x_1, x_2, \ldots, x_n\}$
    and a finite set of clauses $C = \{c_1, c_2, \ldots, c_m\}$, such
    that each clause is logic OR of 3 literals, where the literals over variable $x_i$ are
    $x^0_i$ ($:= x_i$) and $x^1_i$ ($:= \overline{x_i}$).  The problem is to find a
    truth assignment to variables in $X$ such that the logic AND of all clauses in
    $C$ is satisfied.

    Let $\phi = (X, C)$ be an instance of 3SAT.  We can define an
    instance of the DAG alignment problem as follows.  We begin
    by defining the two DAGs used in the alignment.
    First, let us define $G_1 = (V_1, E_1)$ where $V_1$ is defined as follows
    \begin{eqnarray*}
        V_1 = \bigcup_{c_i \in C} (x^{p1}_j, i) \cup (x^{p2}_k, i) \cup (x^{p3}_l, i)
        \mbox{, where } \\ c_i = (x^{p1}_j, x^{p2}_k, x^{p3}_l) \mbox{ and } p1, p2, p3 \in \{0, 1\}
        \mbox{ and } j,k,l \leq n.
    \end{eqnarray*}
    \noindent We define the set of edges $E_1$ by creating directed
    edges over the vertices of $V_1$ as $((x^0_j, i), (x^1_j,
    t))$ for all $j \leq n$ and $i, t \leq m$.

    Now, let us define a second DAG, $G_2 = (V_2, E_2)$.  First, we define $V_2$ as
    \begin{eqnarray*}
        V_2 = \{\{y_1, z_1, y_2, z_2, \ldots y_n, z_n\} \times \{1, 2, \ldots, m\}\} \\ \bigcup
        \{\{c_1, c_2, \ldots, c_m\} \times \{1, 2\}\}.
    \end{eqnarray*}
    \noindent Intuition behind this definition is $y_i$ corresponds
    to $x_i$ and $z_i$ corresponds to $\overline{x_i}$.

    We define $E_2$ by creating directed edges $((z_j, i), (y_j,t))$,
    $((y_j,t), (c_i, 1))$ and $((y_j,t), (c_i, 2))$ for all $j \leq n$ and $i, t \leq m.$

    We now have two DAGs, $G_1$ and $G_2$.  We must define the set $\beta$,
    which describes the possible matches between the two DAGs, and the
    related weights.  For every vertex, $(x^0_j, i)$ or $(x^1_j,
    i)$, map this vertex to its corresponding vertex in $V_2$
    with weight equal to one and add this to $\beta$.  That is, $(x^{0}_j,
    i) \in V_1$ maps to $(y_j, i) \in V_2$
    and $(x^{1}_k, t) \in V_1$ maps to $(z_k,t) \in V_2$, and so forth.
    Also, for each vertex $(x^p_j, i) \in V_1$, create mappings $((x^p_j, i),
    (c_i,1))$ and $((x^p_j, i), (c_i, 2))$ both with weight equal to
    one and add this to $\beta$.  Let the total weight and the total number of
    vertices for the matching be $3m$.

    We now show that the DAG alignment problem, as described
    above, has a matching satisfying the hierarchical mapping
    constraint, if and only if $\phi$ is satisfiable.

    ($\Rightarrow$) Assume $\phi$ is satisfiable.  For each clause $c_i$, choose a
    single literal $x^p_j$.  If variable $x_j \in X$ is true and $p = 0$ or
    $x_j \in X$ is false and $p = 1$, then
    include edge $((x^0_j, i), (y_j, i))$ in the matching $M$.  Also, for
    any clause $c_t$ with $x^1_j$ include edge $((x^1_j, t), (c_t,
    1))$ if vertex $(c_t, 1)$ is not in the matching, otherwise include
    edge $((x^1_j, t), (c_t, 2))$.  Similarly, if variable $x_j \in X$ is
    true and $p = 1$ or $x_j \in X$ is false and $p = 0$, then include
    $((x^1_j, i), (z_j, i))$ in the matching.  Also, for
    any clause $c_t$ with $x^0_j$, include edge $((x^0_j, t), (c_t,
    1))$ if vertex $(c_t, 1)$ is not in the matching, otherwise include
    edge $((x^0_j, t), (c_t, 2))$.  Thus, $M$ exactly maps all vertices
    in $G_1$ to vertices in $G_2$.  There are
    $3m$ vertices in $V_1$, so $|M| = 3m$.  Also, since the
    weight of each edge is one, $w(M) = 3m$.  Finally, since both $x^0_j$ and
    $x^1_j$ cannot be true, both edges $((x^0_j,
    i), (y_j, i))$ and $((x^1_j, i), (z_j, i))$
    cannot be in $M$, therefore the hierarchical constraint is
    satisfied.

    ($\Leftarrow$) Let $M$ be a solution to the DAG alignment problem.  The
    truth value of any variable $x_j$ is assigned as follows.  If, for
    any clause $c_i$ with literal $x^0_j$, there exists an edge
    $((x^0_j, i), (y_j, i))$ from $G_1$ to $G_2$, then let $x_j$ be true.
    Similarly, if there exists an edge $((x^1_j, i), (z_j, i))$ from
    $G_1$ to $G_2$, then let $x_j$ be
    false.  Since in $G_1$, every vertex $(x^0_j,
    i)$ has an edge to every $(x^1_j, t)$, and in $G_2$ every vertex $(z_j, i)$ has an
    edge to every $(y_j, t)$, $M$ cannot contain edges $((x^0_j,
    i), (y_j, i))$ and $((x^1_j, i), (z_j, i))$,
    otherwise the hierarchical constraint would be violated.  Thus, $x^0_j$ or
    $x^1_j$ is true, but never both.  Also,
    since any false literal in a clause $c_i$ is mapped to a vertex
    $(c_i, 1)$ or $(c_i, 2)$, at most 2 vertices in any clause can be
    false.  Thus, $\phi$ is satisfied.
    \end{proof}

\section{Upper bounds on approximating weighted DAG alignment}

Since weighted DAG alignment belongs to the class of NP--complete
problems, it is unlikely that we will find a polynomial time
solution to the problem.  Thus, we must rely on an approximation
scheme for computing alignments.

In this section, we introduce two polynomial time reductions of
the DAG alignment problem to other known NP--complete
problems and use these to provide upper bounds for approximating the
weighted DAG alignment problem.  The quality of the approximation is
given as the ratio between the size of the maximum weighted DAG
alignment and the approximation found.  The ratio in
the worst--case scenario defines the {\it performance guarantee} of
the algorithm.

We begin by reducing the DAG alignment problem to Weighted
Independent Set (WIS).  In the Independent Set problem,
we are given a graph $G = (V, E)$, and we wish to find the largest
subset $S \subseteq V$, such that no two vertices in $S$ are
connected by an edge in $E$.  In the weighted version of this
problem, each node, $v_i \in V$, has an associated weight $w_i$,
and we wish to find the maximum weighted independent set.

H\aa{}stad \cite{hastad99clique} showed that Independent Set is hard to approximate within $n^{1 - \epsilon}$, for $\epsilon > 0$, unless NP--hard 
problems have randomized polynomial time solutions.  In \cite{boppana90approximating}, Boppana and Halld\'{o}rsson introduced the Ramsey algorithm for 
solving WIS.  The algorithm is an extension of the naive greedy approach, where in the greedy approach a vertex $v$ is arbitrarily selected from the 
graph and added to the independent set, all adjacent vertices are removed, and this process is continued until all vertices are exhausted.  The 
obvious problem with this solution is that the adjacencies are ignored.  The first extension to this process is to consider not only the vertex $v$, 
but also the neighbors of $v$.  The algorithm recurses by first considering $v$ as part of the independent set, and then $v$ not in the independent 
set, and selecting the better of the two results.  This algorithm performs well provided the maximum {\it Clique} size is small.  Boppana and 
Halld\'{o}rsson further extended this algorithm by first removing the maximum set of disjoint $k$--cliques, and then apply the Ramsey algorithm to 
compute the independent set on this modified graph.  From this, they were able to prove that the algorithm had a performance guarantee of $O(n/\log^2 
n)$, where $n$ is the number of vertices in the graph.

The following shows that any instance of the DAG alignment
problem can be reduced, in polynomial time, to an instance of
WIS.  This reduction will
allow us to use approximation strategies for Independent Set to
find approximate solutions to the DAG alignment
problem.

\begin{theorem}
The ontology alignment problem can be approximated within
$O(m/log^2 m)$ where $m = |\beta|$.
\end{theorem}

\begin{proof}
    Consider an instance of the
    DAG alignment problem, defined by graphs $G_1 = (V_1, E_1)$
    and $G_2 = (V_2, E_2)$, and the set of edges
    $\beta$.  We define an instance of WIS, by constructing a
    graph $G = (V, E)$ as follows.  For each edge $e = (a, b, w_e) \in
    \beta$, construct a corresponding $v \in V$, and let
    the weight of vertex $v$ be $w := w_e$.  Next, let $E = \{(v_i, v_j) | e_j \in
    conf(e_i) \mbox{ and } e_i, e_j \in \beta \}$.

    Now, we claim that a solution to WIS, defined over graph
    $G$, corresponds to a solution to the DAG alignment
    problem.  We construct this solution as follows.  Let $S$ be
    our solution to WIS.  Then, for each $v_i \in S$, add the
    edge from $\beta$ that corresponds to $v_i$, to our DAG
    alignment solution $M$.  This precisely constructs a
    valid DAG alignment, since each $v_i \in S$ cannot be
    connected to any other $v_j \in S$, which implies that for edges
    $e_i, e_j \in M$, $e_i \not\in conf(e_j)$.  Since no edges
    in $M$ conflict, this must be a valid solution.

    WIS can be approximated within $O(n/log^2 n)$, where $n$ is the number of
    vertices in the graph.  In our
    reduction, $n$ corresponds to $|\beta|$, by letting $m =
    |\beta|$, we achieve an approximation of $O(m/log^2 m)$.
\end{proof}

Next, we improve this bound via a reduction to the Weighted Set
Packing (WSP) problem.  In WSP, we have a set $S$ of $m$ base
elements, and a collection $\mathcal{U} = \{U_1, U_2, \ldots,
U_n\}$ of weighted subsets of $S$.  We want to find a
subcollection $\mathcal{U}' \subseteq \mathcal{U}$ of disjoint
sets of maximum total weight.

In \cite{us99approximation}, an approximation guarantee of $\sqrt{m}$, where $m = |S|$ is given for WSP.  The algorithm is based on a variant of the 
greedy algorithm for solving the non-weighted version introduced in \cite{halldorsson99independent}.  In the following theorem, we show that any 
instance of the DAG alignment problem can be reduced to WSP in polynomial time, and that a solution to WSP corresponds to a solution of the DAG 
alignment problem.

\begin{theorem}
The DAG alignment problem can be approximated within
$\sqrt{m}$ where $m = |\beta|$.
\end{theorem}

\begin{proof}
    Consider an instance of the
    DAG alignment problem, defined by graphs $G_1 = (V_1, E_1)$
    and $G_2 = (V_2, E_2)$, and the set of edges
    $\beta$.  We define an instance of WSP, by constructing $S$
    and the collection $\mathcal{U}$ as follows.

    We let our $m$ base
    elements be the edges specified by $\beta$, thus our set $S
    = \beta$.  We construct the collection $\mathcal{U}$, by defining
    subsets $U_i$ for all $e_i \in \beta$ as $U_i = \{ \{e_i\} \bigcup
    conf(e_i)\}$.  Let the weight of $U_i$ be equal to $w_{e_i}$.
    We now claim that any solution to WSP, $\mathcal{U}'$,
    corresponds to a solution the DAG alignment problem.

    We can see this by considering any $\mathcal{U}'$.  We construct a
    solution to the DAG alignment problem by taking each $U_i
    \in \mathcal{U}'$, and adding edge $e_i \in \beta$ to our
    ontology alignment solution $M$.  This is a valid matching
    because every $U_i \in \mathcal{U}'$ is disjoint, which
    implies that for each $e_i \in M$ and $e_j \in M$, $e_i
    \not\in conf(e_j)$, so no edges in $M$ conflict.

    Since a solution to WSP yields a solution to the
    DAG alignment problem, approximations of WSP correspond to
    approximations of the DAG alignment problem.  Hence, we can
    approximate the DAG alignment problem within
    $\sqrt{m}$, where $m = |\beta|$.
\end{proof}

\section{Polynomial-time algorithms}
In this section we study certain types/classes of graphs with respect to their DAG
alignment problem solution complexity. In particular, we show that the
DAG alignment problem for trees has a polynomial time solution.
In this work, we naturally define trees to be those directed trees with all edges 
directed away from a particular vertex called the root. In this section we show 
that any two such trees can be aligned in polynomial time. Furthermore, a 
{\em chain} $C_n$ is defined as a DAG with $n$ vertices $v_1,v_2,...,v_n$ and
directed edges $(v_1,v_2),\;(v_2,v_3),...\;,(v_{n-1},v_n)$.

\begin{theorem}\label{hoyer}
Any two trees can be aligned in polynomial time.
\end{theorem}

\begin{proof}
We first describe the data structure used in our algorithm, and then explain how it can be used 
to achieve a polynomial time algorithm that aligns two trees. Our algorithm 
is a form of bottom-up approach that applies weighted bipartite matching at each of $n \times k$ 
iterations it makes.

Suppose we have two trees $T^1=(V^1,E^1)$ and $T^2=(V^2,E^2)$ 
with $n$ and $k$ vertices correspondingly that need to be aligned. Create an
array with $n \times k$ empty cells $C(i,j)\;(i=1..n,j=1..k)$ that 
contain real numbers whose values will be assigned during 
the algorithm and will hold values for best alignment of the subtree of $T^1$ 
with root in $v^1_i$ with the subtree of $T^2$ with root in $v^2_j$. This array 
is complemented by an equal size array $M(i,j)\;(i=1..n,j=1..k)$ that contains 
the actual matchings used for the assigned values of $C(i,j)$. 
We will further describe how to assign values to $C(i,j)$, 
sometimes omitting the discussion of updates to $M(i,j)$.
Our algorithm terminates when $C(n,k)$ gets assigned a value. Once this is 
done, the value stored in $C(n,k)$ equals to the maximum weight alignment 
and $M(n,k)$ contains the best matching.

We next describe the total order on the set of vertices of both trees and order
cells $C(i,j)$. Consider tree $T^1$. Suppose its depth is $d$. Name all vertices 
at level $d$, $v^1_1$ through $v^1_{d_1}$ (for instance, name these vertices in 
the left to right order assuming the tree is drawn on paper with no edge 
intersections) for appropriate value of $d_1$. 
Next, name all depth $d-1$ vertices, $v^1_{d_1+1}$ through $v^1_{d_2}$, for 
appropriate value $d_2$. Continue this operation until all vertices are named. 
Vertex $v^1_n$ is thus the root of tree $T^1$. Apply the same method to 
enumerate vertices in tree $T^2$. Cells $C(i,j)$ are ordered lexicographically,
{\em e.g.} $C(1,1) \prec C(1,2) \prec ... \prec C(1,k) \prec C(2,1) \prec C(2,2) 
\prec ... \prec C(n,k)$. We fill values $C(i,j)$ (and keep track of alignment 
made by updating $M(i,j)$) in this order.

$C(1,1)$ is easy to find, because it is equal to the weight of edge $e=(v^1_1,v^2_1,w_e)$
of the matching problem. To find the value of $C(i,j)$ (and update $M(i,j)$) consider the 
following cases (with $C(i,j)$ taking the maximal value among those found in each of the 
cases below):
\begin{enumerate}
\item $v^1_i$ does not get mapped anywhere. In this case, 
$C(i,j)=\max\{C(v^1_t,v^2_j)|v^1_t=child(v^1_i)\}$. Each such $C(v^1_t,v^2_j) \prec C(i,j)$
and thus the maximum, is well defined and can be calculated. The computational cost of this calculation 
is the number of children of $v^1_i$, {\em i.e.} no more than $n$.
\item $v^1_i$ is mapped to $v^2_k \in desc(v^2_j)$, and hence $k<j$. In this case, 
$C(i,j)=w(v^1_i,v^2_k) + S$, where $S$ is the answer to the following weighted bipartite 
matching problem. Assuming $v^1_i$ has children $child^1_1,child^1_2,...,child^1_{ic}$ and 
$v^2_k$ has children $child^2_1,child^2_2,...,child^2_{kc}$, the maximum bipartite matching 
problem whose solution is the number $S$ we are interested in is defined for the 
complete bipartite graph with vertices 
$child^1_1,child^1_2,...,child^1_{ic},child^2_1,child^2_2,$ $...,child^2_{kc}$ and edges 
with weights $C(child^1_s,child^2_t)|_{s=1..ic,t=1..kc}.$ Note that all such weights 
are known and thus the problem is well defined. The solution to the maximum weighted 
bipartite matching can be found in polynomial time, and the number of times we 
call for a solution is limited by the number of descendants
of $v^2_j$, which is never more than $k$ (the number of vertices in tree $T^2$). Thus,
this step can be completed in polynomial time.
\end{enumerate}

The number of different $C(i,j)$ is polynomial, and the amount of work required to fill 
in each value is polynomial. Thus, our algorithm is polytime. For  
two trees with $n$ vertices each, the complexity of our algorithm is $O(n^6)$: there 
are $n^2$ numbers $C(i,j)$ to calculate, and calculation of each requires (item 2) 
at most $n \times n^3$ operations assuming the Hungarian algorithm \cite{kuhn55hungarian} for weighted bipartite 
matching is used.
\end{proof}

It appears that the complexity of the DAG alignment problem moves from P to NP--complete
in transition from trees to DAGs. The part of the above proof that works for trees and breaks
for DAGs is the ability to establish an order on the numbers $C(i,j)$ such that once a particular 
$C(i,j)$ has been calculated it never needs to get updated. 

The described polynomial time algorithm requires $O(n^6)$ runtime to align two trees. 
However, for some simpler types of trees the polynomial time complexity can be reduced through 
considering simplified and modified versions of the above algorithm. A detailed description of 
such algorithms is out of scope for this paper. However,
we would like to mention that two chains (with $n$ vertices each) can be aligned with a cost of $O(n^3)$ 
and two complete binary trees with the cost of $O(\frac{n^4}{\log(n)})$.

\section{Conclusions}
We introduced a new type of weighted matching problem called the weighted hierarchical DAG alignment problem.  We formalized this problem, showed that 
it is NP--complete, proved several upper bounds for approximating solutions to the problem, and finally introduced algorithms for solving different 
classes of the problem.  This problem developed through our research on ontology alignment, however, it relates to many different applications, 
including, but not limited to, UML diagram comparison, SVG document comparison, and file/folder mapping. Our results show that, in particular, 
file/folder mapping problem can be solved in polynomial time, since the underlying data structure is a tree.

In the future, we plan to find other classes of DAGs that can be aligned faster than with an exponential time algorithm, work on designing efficient 
heuristics, and finally apply some of these ideas to the problem of aligning ontologies.

With ontologies, the problem becomes even more complex because they can contain errors in their specification, meaning that in some circumstances the 
hierarchical constraint must be relaxed.  Moreover, this is likely the case with other applications of the problem.  Thus, it may also be an 
interesting problem to investigate approximate solutions that are allowed to contain a small number of edge conflicts, which will accommodate for some 
human error in an ontology specification.

\section*{Acknowledgements}
We wish to acknowledge Prof. P. H{\o{}}yer from University of Calgary for his help in proving Theorem \ref{hoyer}.


This work was supported in part by National Center for Biomedical Ontology, under roadmap-initiative grant U54 HG004028 from the National Institutes 
of Health, and by the PDF grant from the Natural Sciences and Engineering Research Council of Canada (NSERC).

\bibliography{refs}

\begin{thebibliography}{10}

\bibitem{almohamad93linear}
H.~A. Almohamad and S.~O. Duffuaa.
\newblock A linear programming approach for the weighted graph matching
  problem.
\newblock {\em IEEE Transactions on Pattern Analysis and Machine Intelligence},
  15(5):522--525, 1993.

\bibitem{boppana90approximating}
R.~Boppana and M.~M. Halld\'{o}rsson.
\newblock Approximating maximum independent sets by excluding subgraphs.
\newblock In J.~R. Gilbert and R.~Karlsson, editors, {\em {SWAT} 90 2nd
  Scandinavian Workshop on Algorithm Theory}, volume 447, pages 13--25, 1990.

\bibitem{buss95bipartite}
S.~Buss and P.~Yianilos.
\newblock A bipartite matching approach to approximate string comparison and
  search.
\newblock Technical report, 1995.

\bibitem{caseau97solving}
Yves Caseau and Francois Laburthe.
\newblock Solving various weighted matching problems with constraints.
\newblock In {\em Principles and Practice of Constraint Programming}, pages
  17--31, 1997.

\bibitem{cs.DS/0604037}
E.~D. Demaine, S.~Mozes, B.~Rossman, and O.~Weimann.
\newblock An o(n\^{}3)-time algorithm for tree edit distance.
\newblock http://arxiv.org/abs/cs.DS/0604037, April 2006.

\bibitem{eshera84imageanalysis}
M.~A. Eshera and K.~S. Fu.
\newblock A graph distance measure for image analysis.
\newblock {\em IEEE Transactions on Systems, Man, and Cybernetics},
  14(3):398--408, 1984.

\bibitem{jrme04state}
J.~Euzenat, T.~Le Bach, J.~Barrasa, P.~Bouquet, J.~DeBo, R.~Dieng-Kuntz,
  M.~Ehrig, M.~Hauswirth, M.~Jarrar, R.~Lara, D.~Maynard, A.~Napoli, G.~Stamou,
  H.~Stuckenschmidt, P.~Shvaiko, S.~Tessaris, S.~Van Acker, and I.~Zaihrayeu.
\newblock State of the art on ontology. deliverable d2.2.3, 2004.

\bibitem{feng05conceptual}
Y.~Feng, R.~L. Goldstone, and V.~Menkov.
\newblock A graph matching algorithm and its application to conceptual system
  translation.
\newblock {\em International Journal on Artificial Intelligence Tools},
  14:77--100, 2005.

\bibitem{galil86weightedmatching}
Z.~Galil, S.~Micali, and H.~Gabow.
\newblock An $o(ev \log v)$ algorithm for finding a maximal weighted matching
  in general graphs.
\newblock {\em SIAM J. Comput.}, 15(1):120--130, 1986.

\bibitem{garey79npcomplete}
M.~R. Garey and D.~S. Johnson.
\newblock {\em Computers and Intractability : A Guide to the Theory of
  NP-Completeness}.
\newblock W. H. Freeman, 1979.

\bibitem{gold96graduated}
S.~Gold and A.~Rangarajan.
\newblock A graduated assignment algorithm for graph matching.
\newblock {\em IEEE Transactions on Pattern Analysis and Machine Intelligence},
  18(4):377--388, 1996.

\bibitem{gruber93transapproach}
T.~R. Gruber.
\newblock A translation approach to portable ontology specifications.
\newblock {\em Knowledge Acquisition}, 5(2):23--28, 1993.

\bibitem{hastad99clique}
J.~H\aa{}stad.
\newblock Clique is hard to approximate within $n^{1 - \epsilon}$.
\newblock {\em Acta Mathematica}, 182:105--142, 1999.

\bibitem{halldorsson99independent}
M.~Halld\'{o}rsson, J.~Kratochvil, and J.~Telle.
\newblock Independent sets with domination constraints.
\newblock {\em Discrete Applied Mathematics}, 99:39--54, 1999.

\bibitem{us99approximation}
M.~M. Halld\'{o}rsson.
\newblock Approximation of weighted independent set and hereditary subset
  problems.
\newblock In {\em Proceedings of COCOON'99}, 1999.

\bibitem{hummel83relaxation}
R.~A. Hummel and S.~W. Zucker.
\newblock On the foundations of relaxation labeling processes.
\newblock {\em IEEE Transactions on Pattern Analysis and Machine Intelligence},
  5(3), 1983.

\bibitem{krcmar94genetic}
M.~Krcmar and A.~Dhawan.
\newblock Application of genetic algorithms in graph matching.
\newblock In {\em International Conference on Neural Networks}, volume~6, pages
  3872--3876, 1994.

\bibitem{kuhn55hungarian}
H.~W. Kuhn.
\newblock The hungarian method for the assignment problem.
\newblock {\em Naval Research Logistics Quaterly}, 2:83--95, 1955.

\bibitem{kuner88neural}
P.~Kuner and B.~Ueberreiter.
\newblock Pattern recognition by graph matching combinatorial versus continuous
  optimization.
\newblock {\em International Journal Pattern Recognition and Artificial
  Intelligence}, 2:527--542, 1988.

\bibitem{niere04uml}
J.~Niere.
\newblock Visualizing differences of uml diagrams with fujaba.
\newblock In {\em Proceedings of the Fujaba Days 2004}, 2004.

\bibitem{noy02prompt}
N.~Noy and M.~Musen.
\newblock The prompt suite: Interactive tools for ontology merging and mapping.
\newblock Technical report, 2002.

\bibitem{noy99smart}
N.~F. Noy and M.~A. Musen.
\newblock An algorithm for merging and aligning ontologies: Automation and tool
  support.
\newblock In {\em Sixteenth National Conference on Artificial Intelligence
  (AAAI-99), Workshop on Ontology Management}, 1999.

\bibitem{rangarajan94lagrangian}
A.~Rangarajan and E.~Mjolsness.
\newblock {A Lagrangian relaxation network for graph matching}.
\newblock In {\em International Conference on Neural Networks}, volume~7, pages
  4629--4634. Inst. Electrical {\&} Electronics Engineers, 1994.

\bibitem{russel95modern}
S.~Russell and P.~Norvig.
\newblock {\em Artificial Intelligence: A Modern Approach}.
\newblock Prentice Hall, Upper Saddle River, New Jersey, 1995.

\bibitem{shapiro81inexactmatching}
L.~Shapiro and R.~Haralick.
\newblock Structural descriptions and inexact matching.
\newblock {\em IEEE Transactions on Pattern Analysis and Machine Intelligence},
  3:504--519, 1981.

\bibitem{tsai83patternrecognition}
W.~H. Tsai and K.~S. Fu.
\newblock Subgraph error-correcting isomorphism for syntactic pattern
  recognition.
\newblock {\em IEEE Transactions on Systems, Man and Cybernetics}, 13:48--62,
  1983.

\bibitem{umeyama88eigen}
S.~Umeyama.
\newblock An eigendecomposition approach to weighted graph matching problems.
\newblock {\em IEEE Transactions on Pattern Analysis and Machine Intelligence},
  10(5):695--703, 1988.

\bibitem{zhang95editing}
K.~Zhang, J.~T.~L. Wang, and D.~Shasha.
\newblock On the editing distance between undirected acyclic graphs and related
  problems.
\newblock In {\em Proceedings of the 6th Annual Symposium on Combinatorial
  Pattern Matching}, volume 937, pages 395--407. Springer-Verlag, 1995.

\end{thebibliography}
\end{document}